\documentclass[runningheads]{llncs}
\usepackage{longtable}
\usepackage{color}
\usepackage{booktabs}
\usepackage{array}
\usepackage[hang,small,bf]{caption}
\usepackage[subrefformat=parens]{subcaption}
\usepackage[T1]{fontenc}
\usepackage{graphicx}
\usepackage{enumitem}
\newlist{questions}{enumerate}{2}
\setlist[questions,1]{label=RQ\arabic*.,ref=RQ\arabic*}
\setlist[questions,2]{label=(\alph*),ref=\thequestionsi(\alph*)}

\begin{document}

\title{Investigating Industry--Academia Collaboration in Artificial Intelligence: PDF-Based Bibliometric Analysis from Leading Conferences}
\titlerunning{Investigating Industry--Academia Collaboration in Artificial Intelligence}
%
\author{Kazuhiro Yamauchi\orcidID{0009-0000-1206-2426} \and \\ Marie Katsurai\orcidID{0000-0003-4899-2427}}
\authorrunning{K. Yamauchi et al.}
%
\institute{Doshisha University, Kyotanabe-shi Kyoto 610-0394, Japan
\email{\{yamauchi23,katsurai\}@mm.doshisha.ac.jp}}
%
%
\maketitle              
\begin{abstract}
This study presents a bibliometric analysis of industry--academia collaboration in artificial intelligence (AI) research, focusing on papers from two major international conferences, AAAI and IJCAI, from 2010 to 2023. Most previous studies have relied on publishers and other databases to analyze bibliographic information. However, these databases have problems, such as missing articles and omitted metadata. Therefore, we adopted a novel approach to extract bibliographic information directly from the article PDFs: we examined 20,549 articles and identified the collaborative papers through a classification process of author affiliation. The analysis explores the temporal evolution of collaboration in AI, highlighting significant changes in collaboration patterns over the past decade. In particular, this study examines the role of key academic and industrial institutions in facilitating these collaborations, focusing on emerging global trends. Additionally, a content analysis using document classification was conducted to examine the type of first author in collaborative research articles and explore the potential differences between collaborative and noncollaborative research articles. The results showed that, in terms of publication, collaborations are mainly led by academia, but their content is not significantly different from that of others. The affiliation metadata are available at \url{https://github.com/mm-doshisha/ICADL2024}.
\footnote{This version of the contribution has been accepted for publication at ICADL 2024, after peer review but is not the version of record and does not reflect post-acceptance improvements, or any corrections.}
\end{abstract}

\section{Introduction}
In the modern world, addressing complex societal issues and scientific challenges, requires a team science approach that integrates diverse perspectives and expertise. Industry--academia collaboration plays a crucial role in bridging theory and practice to achieve a more direct societal impact. Perkmann et al. \cite{10.1093/icc/dtp015} demonstrated that such collaborations enable two-way knowledge transfer: Industry gains access to the latest academic knowledge, whereas academia learns about real-world challenges, potentially inspiring new research directions. This mutual interaction is particularly important in the rapidly evolving field of artificial intelligence (AI), which is having an increasing impact on society. Addressing the challenges associated with the implementation of AI technologies requires an effective combination of cutting-edge research and practical needs. This integration is expected to further drive innovation in the field. To gain insights into the current state and future directions of AI research and development, it is essential to analyze trends and collaboration patterns in the field.
One widely recognized method for understanding industry trends and research is bibliometric analysis. This approach provides valuable insights by examining publication patterns or collaboration structures within a given field \cite{donthu2021conduct}. Therefore, applying bibliometric analysis to the AI field is an important task that can yield significant benefits. The insights gained from such an analysis are valuable not only to the academic community but also to policymakers and industry stakeholders. \par
Detecting industry--academia collaboration papers requires the identification of author affiliation types. Previous bibliometric studies focusing on author affiliations \cite{farber2024analyzing,Gao2024-ul,Hu2020-bx,Iqbal2019-fb,Jee2022-gx} have used publisher databases such as Scopus and Web of Science (WoS) and open data sources such as Microsoft Academic Graph (MAG) and Semantic Scholar (S2) as information sources. However, these databases do not necessarily contain all article information, nor do they always include author affiliation information as metadata, which is essential for analyzing industry--academic collaboration. This can limit the coverage and accuracy of the analysis. In particular, papers from top-tier conferences in the fields of AI---AAAI and IJCAI---are not included in these databases or are not accurately recorded. Therefore, in this study, we collected PDFs of papers from these major AI conferences and directly extracted the bibliographic information. Then, by focusing on the authors' affiliations within the extracted bibliographic information, we determined whether they were from academia or industry. This has allowed us to identify papers that involve academia and industry, which we consider industry--academia collaborative papers. A bibliometric analysis of these papers is then performed to clarify the status of collaboration in the field of AI. Our research questions for the field of AI are as follows:
\begin{questions}[itemsep=1ex, leftmargin=1cm]
\item Are industry--academia collaborative papers increasing annually?
\item Which research institutions are most actively engaged in industry--academia collaboration?
\item Is academia or industry leading industry--academia collaboration?
\item Are there differences in content between industry--academia collaborative papers and other papers?
\end{questions}

\section{Related Works}
\subsection{Bibliometric Analysis of Industry--Academia Collaboration}
Many studies have analyzed the benefits of collaboration between universities and industry, exploring best practices for such partnerships. However, there are fewer examples of bibliometric analyses that have focused on papers produced through industry--academia joint efforts. Existing works have focused on trends in particular countries \cite{Abramo2009-kg,Calvert2003-za,Zhou2016-ay} or research areas \cite{BUTCHER20051273}.
Calvert et al. \cite{Calvert2003-za} examined several aspects of university--industry collaboration by analyzing papers coauthored by universities and companies in the UK from 1981 to 2000; this included an analysis of coauthorship patterns, publication years, fields of study, and characteristics of the participating universities and companies. Through this bibliometric analysis, they observed the quantitative growth of papers resulting from university--industry partnerships over time and trends in different academic fields where such collaborations were prevalent. They further identified the top institutions that are the most active in industry--academia partnerships, examining the extent to which foreign firms are involved in these research alliances. 
Butcher et al. \cite{BUTCHER20051273} conducted an analysis of industry--academic collaborations on a specific research topic, that is, the application of membranes in water treatment. The results showed that an increase in industry--academia joint efforts coincided with a sharp increase in patent applications after 1995, suggesting that such cooperation may have contributed to technological innovation.
Using bibliometric methods, such as field-specific mapping and quality analysis, Abramo et al. \cite{Abramo2009-kg} analyzed university--industry research collaborations in Italy. The study revealed that joint projects were most common in medicine and chemistry and that academic researchers collaborating with industry demonstrated higher search productivity. 
Using data from the Center for Science and Technology Studies Leiden Ranking, Zhou et al. \cite{Zhou2016-ay} analyzed university--industry coauthored publications from China and the United States indexed in WoS between 2009 and 2012; they examined publication productivity, collaboration intensity, and geographic patterns of cooperation for universities in both countries. The results showed that universities in the United States were more active in industry--academia partnerships than Chinese universities, with higher publication productivity and collaboration intensity. \par
Country-specific and sector-specific analyses of industry--academia collaboration can provide detailed information on trends particular to the area, thus contributing to effective strategic planning and policy-making. Furthermore, these analyses assist in the optimal distribution of resources by elucidating the relevance of technological innovations and the attributes of disparate sectors. However, there are no examples of collaboration analyses focused on the field of AI, which is developing rapidly and has a significant impact on society. Therefore, the present study aims to clarify the current state and characteristics of industry--academia collaborations in AI.

\subsection{Bibliometric Analysis Focusing on Companies in the AI Field}
Although several bibliometric analyses have been conducted in response to the growing number of AI-related papers, few have focused on companies that have contributed to the development of the field. For example, Helene et al. \cite{doi/10.2760/472704} showed that information and communication technology companies are leading the development of AI technology by analyzing AI-related patents, trademarks, and papers; they also reported that approximately half of AI-related patents are held by companies in the computer and electronics sectors. In addition, Jee et al. \cite{Jee2022-gx} analyzed AI-related papers in Scopus and MAG, reporting that papers involving companies, especially those by company researchers alone, have a greater impact than other papers in terms of several measures, such as the h-index. Jee et al. \cite{Jee2022-gx} also showed that papers resulting from industry--academia collaborations are more likely to produce novel and important knowledge compared to other papers.
Finally, F\"arber et al. \cite{farber2024analyzing} investigated the inﬂuence of companies in AI research, utilizing the MAG and Altmetric.com database\footnote{https://www.altmetric.com/} as the data sources. The findings revealed that papers involving companies had significantly more citations and greater online attention than papers from academic institutions alone. Additionally, companies that published AI-related papers particularly focused on fields inﬂuenced by deep learning, such as computer vision and natural language processing. The study demonstrated that Big Tech entities, including Google, Facebook/Meta, and Microsoft, made notable contributions to AI research, as evidenced by their high citation and altmetric scores. Although several bibliometric analyses have focused on companies in the field of AI and some mention coauthorship between universities and companies, no papers have solely focused on industry--academia partnerships in their analysis and discussion. Moreover, these studies rely on publisher databases or open data. Although it is widely known that cutting-edge results in AI are often presented at international conferences \cite{Gao2024-ul}, existing databases contain mainly journal articles and lack bibliographic information on these conferences. As a result, our study has employed a bibliometric analysis focusing on industry--academia collaboration by collecting PDFs of papers from international conference websites and extracting bibliographic information directly from them.
\section{Dataset Construction}
We selected AAAI and IJCAI, top-tier conferences whose names include the word ``AI,'' which our analysis targets. We obtained PDFs in conference proceedings by scraping the official websites from 2010 to 2023. Notably, especially for AAAI, the proceedings page also includes papers from other conferences in the same association (e.g., IAAI, EAAI), but we collected only papers from the AAAI Conference on Artificial Intelligence. A total of 20,549 papers were collected, with 12,517 from AAAI and 8,032 from IJCAI. 
The procedure for preparing paper metadata involves several steps. First, bibliographic information was extracted directly from the PDF files (see Section 3.1). We subsequently normalized the authors' affiliation information from the extracted data and classified it into academic or company data (see Section 3.2). Through this process, papers coauthored by researchers affiliated with academia and those affiliated with companies were identified and considered products of industry--academia collaboration. This methodical approach enabled the systematic identification and analysis of collaborative research outputs between the academic and industrial sectors. 
\begin{figure}[t]
\includegraphics[width=0.4\textwidth,bb=9 9 358 434]{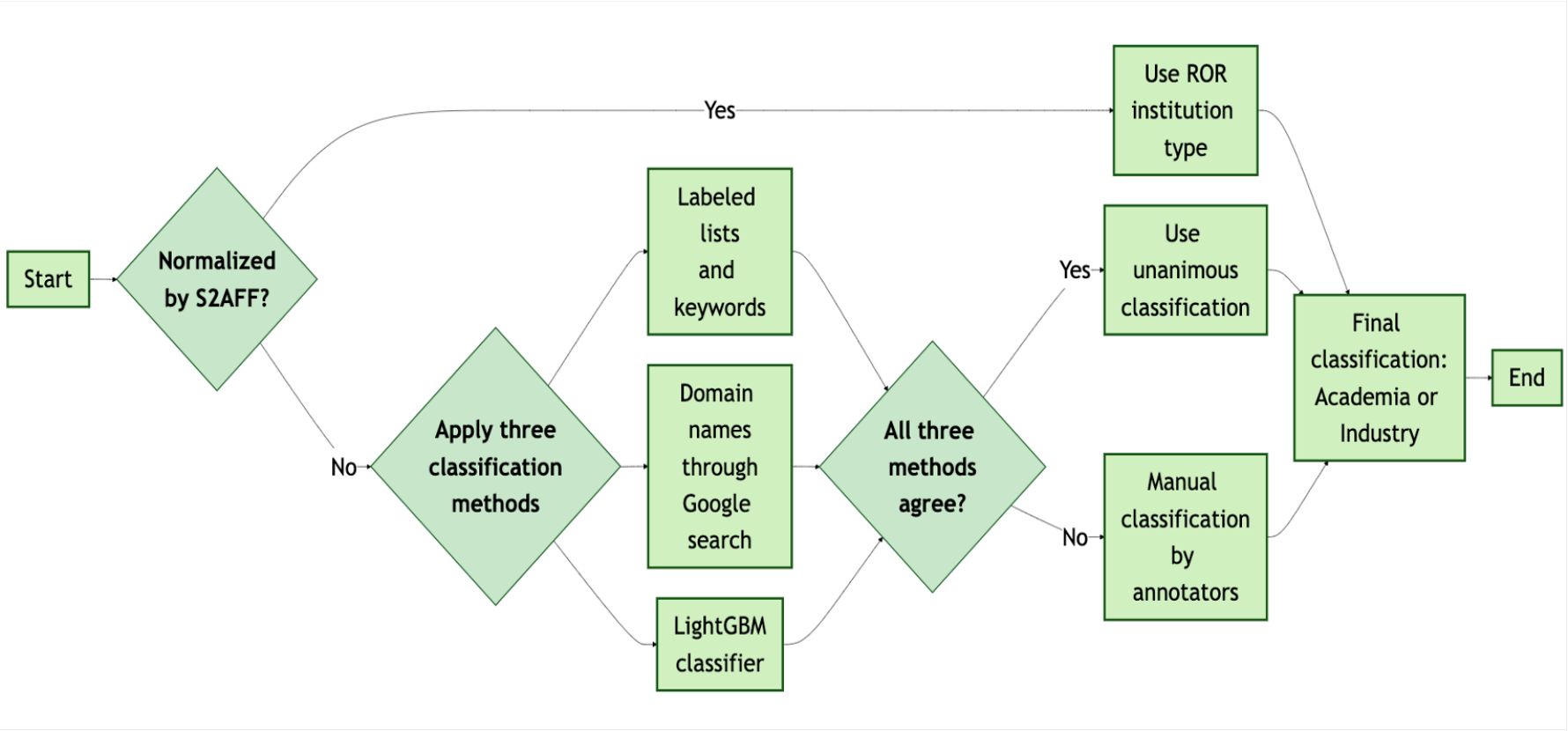}
\caption{The flowchart of the classification of whether the institution is academic or industrial.}
\label{fig:flowchart}
\end{figure}

\subsection{Extracting Author Fields from Scientific PDF Documents}

To extract the author fields from the collected PDFs, we used GROBID\footnote{https://grobid.readthedocs.io/en/latest/}, which can extract information such as titles, abstracts, author names, and affiliations in XML format. We applied GROBID to the collection of paper PDFs and extracted all strings identified as author affiliations. Although GROBID is known to have the highest performance among existing tools \cite{tkaczyk2018machine}, it sometimes produces omissions and errors. For example, strings such as the following were obtained:
``Department of Statistics and Actuarial Science University of Waterloo, 200 University Avenue West Waterloo, Ontario, Canada.''
This string includes not only the name of the affiliated institution but also the address. Additionally, there might be writing variations for a single institution, such as the following:
\begin{itemize}
\item ``EECS Department, University of California, Berkeley Berkeley, CA''       
\item ``Electrical Engineering and Computer Sciences Department, University of California, Berkeley''
\end{itemize}
To resolve these notation inconsistencies, we used S2AFF \cite{kinney2023semantic}, a model that normalizes a given affiliation text using research institution names registered in the Research Organization Registry (ROR)\footnote{https://ror.org}.
Specifically, the named entity recognition model based on Transformer \cite{vaswani2017attention} was used to extract the institution name from the raw affiliation string as enumerated above.
It matches with research institution names registered in the ROR and outputs the most appropriate name from among the candidates. This allows the above strings to be converted to ``University of California, Berkeley.'' We normalized only when the matching score is higher than threshold\footnote{In this study, we set the threshold to 0.9.}, and left the original string unchanged when the score is below the threshold. As a result, 22,408 of the 25,581 institutional strings, excluding duplicates, exceeded the  specified threshold and were normalized. The removal of duplicates from the 22,408 strings resulted in 2,648. The number of strings that were not normalized was 3,173.
\subsection{Classification of Whether the Institution Is Academic or Industrial}

To identify industry--academia collaborative papers, it is necessary to know whether the author's affiliation is with academia or industry. However, there is no simple and automatic way to classify all research institutions as academic or industry. Therefore, we constructed a method based on a linkage with the ROR, which assigns an institution type to each institution, so that a string matched by S2AFF can be automatically classified. Figure \ref{fig:flowchart} shows a flowchart of the constructed classification method. There are the following eight institution types in total: Education, Healthcare, Company, Archive, Nonprofit, Government, Facility, and Other. Of these, only Company is considered an industry. \par
For the 3,173 institutions not normalized by S2AFF, we used a hybrid approach based on both automatic and manual inspections. Specifically, we prepared the following three binary classification methods: (i) classification via labeled lists and keywords, (ii) classification via domain names via Google search, and (iii) classification via LightGBM \cite{ke2017lightgbm}. If the judgments of these three methods were consistent, we directly used the results for confident classification. Otherwise, human annotators manually classified each affiliation string into academic or industry. The details of each classification method are explained below.
\subsubsection{Classification using labeled lists and keywords.}
We used a list of research institution names labeled as either academia or industry, which was provided in the literature \cite{abuwala2023should}. Given that this list does not cover all institutions, we established separate classification rules for institutions not on the list. Specifically, institutions containing any of the seven keywords---Universi, Academ, School, Polytech, Department, Univ. and Dept.---were classified as academic, and the rest were classified as industry.
\subsubsection{Classification using domain names through Google search.}
We removed postal codes, country names, and regional abbreviations from the institution strings.  Specifically, after splitting the string by commas, we removed substrings containing numbers or two to three consecutive capital letters. After applying this simple filtering, we used the resulting string as a query for the Google search engine. If any of the top three search results contained URLs with ``.edu,'' ``.ac,'' or ``.gov,'' we classified them as academic or industry.
\subsubsection{Classification using LightGBM.}
In addition to the rule-based keyword processing described above, there may be words characteristic of academia and companies. In the current study, we utilized pairs of strings and institution types present in the ROR as ground truths and trained a binary classifier to identify characteristic words. Specifically, 36,897 institution strings were collected from AAAI, IJCAI, and other conferences in computer science (e.g., NeurIPS, ICML, ICLR) and normalized by S2AFF. Because 31,785 normalized institutions are registered in the ROR, these institutions were used to construct the training dataset. Of these, 30,581 were drawn from academic sources, whereas 1,204 were derived from industry sources. The data were split into three partitions, with 80\% used for training, 10\% for validation, and 10\% for testing.  A word2vec~\cite{Mikolov2013-ef} model pretrained on Wikipedia\footnote{https://wikipedia2vec.github.io/wikipedia2vec/pretrained/} was used to calculate 500-dimensional textual features. These text features were then input into a LightGBM classifier, resulting in an ROC-AUC of 0.9494.
\begin{figure}[t]
\includegraphics[width=0.2\textwidth,bb=9 9 358 434]{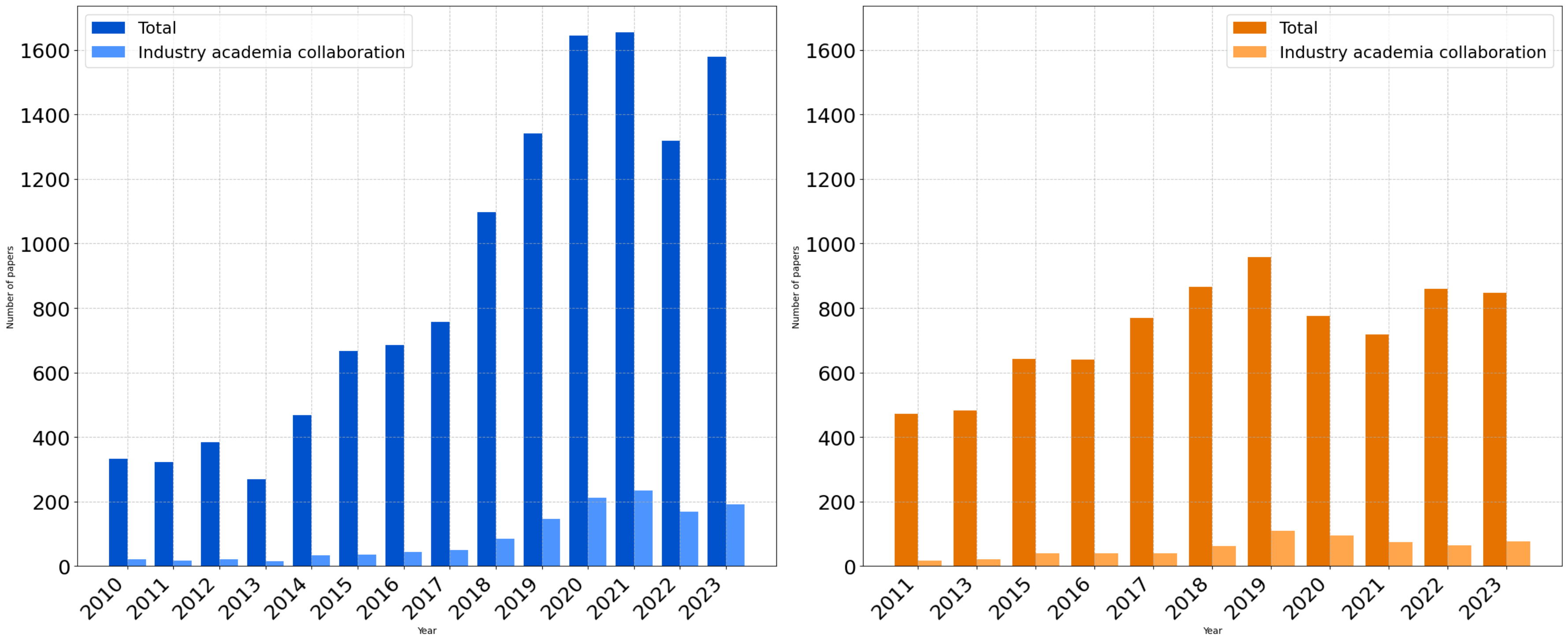}
\caption{Time series changes in the number of papers and industry--academia collaborative papers for each conference.}
\label{fig:basic}
\end{figure}
\subsubsection{Manual classification.}
From the above three classifiers' unanimous votes, 3,970 research institutions were automatically identified as academic or industrial institutions. For the remaining 1,842 institutions with split votes, four annotators, including the first author, visually inspected the author affiliation on the original PDFs of the corresponding articles and searched for the institution's information via web search engines to determine whether it was academia or industry. Manual classification took approximately 35 hours.\\
\par

After identifying all institution types, we counted the number of academic and industry affiliations for each paper. If both were greater than zero, the paper was considered as being an industry--academia collaborative paper. After these classifications, 1,919 papers were determined to be collaborative papers. We intend to make the resulting dataset publicly available.

\section{Statistical Analysis}

\begin{figure}[t]
\includegraphics[width=0.3\textwidth,bb=9 9 358 434]{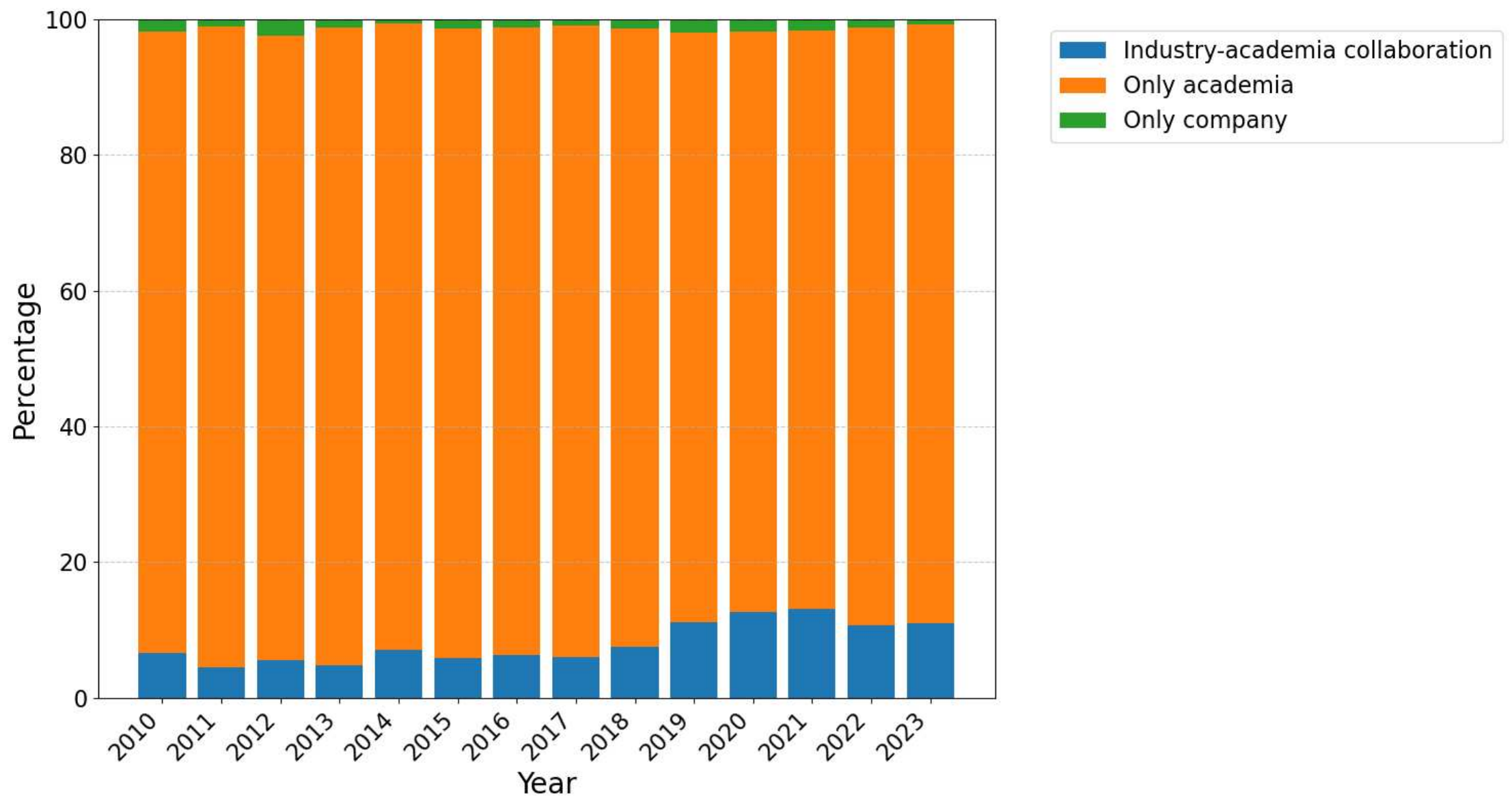}
\caption{The proportion of papers authored solely by academia, solely by industry, and through industry--academia collaborations.}
\label{fig:basic_proportion}
\end{figure}

\subsection{Basic Statistics}

Figure \ref{fig:basic} shows the number of papers and industry--academia collaborative papers by year for each conference. The number of papers on AAAI remained almost unchanged from 2010 to 2013, increased slightly from 2014 to 2017, showed an increasing trend from 2018 to 2021, and has been decreasing since then. For IJCAI, the number of papers remained unchanged in 2011 and 2013, increased slightly in 2015, remained unchanged until 2017, and showed an increasing trend in 2018 and 2019 but has been decreasing since then, with only a slight increase in 2023 compared with 2022. \par
Figure \ref{fig:basic_proportion} shows the change over time in the proportions of papers authored solely by academia, solely by industry, and by collaboration. The proportion of collaborative papers remained unchanged from 2010 to 2017 and showed an increasing trend from 2018 to 2020 and 2021. Since then, the measure has been on a downward trend. From this perspective, the trend seems to have started in approximately 2017 and 2018. In 2022 and 2023, the trend seems to have cooled compared with the wave in approximately 2020. \par
The results indicate that there has been a notable expansion in the partnership between industry and academia in the field of AI, particularly compared with the situation around 2010. This is evident from the increasing trend in both the number and proportion of the papers published since 2017 and 2018. Although Microsoft Research Asia was the institution with the greatest number of collaborations in 2017, the Alibaba Group took the top position in 2019. This implies that, although Big Tech companies have been leading AI research \cite{farber2024analyzing}, companies such as Alibaba Group and Tencent may also be starting to focus on this collaboration.

\subsection{Number of Industry--Academia Collaborative Papers by Research Institution}
\begin{table}[t]
\centering
\caption{Top 10 academic and industry institutions by number of papers.}
\begin{tabular}{l|c||l|c}
\noalign{\hrule height 1.0pt}
\multicolumn{2}{c||}{\textbf{Top 10 academic institutions}} & \multicolumn{2}{c}{\textbf{Top 10 industry institutions}} \\
\noalign{\hrule height 1.0pt}
Institution & Papers & Institution & Papers \\
\noalign{\hrule height 1.0pt}
Zhejiang University & 321 & Microsoft Research Asia & 612 \\
Tsinghua University & 303 & Alibaba Group & 544 \\
Peking University & 265 & Microsoft Research/Microsoft & 407 \\
University of Science and Technology of China & 234 &  Tencent & 203\\
University of Chinese Academy of Sciences & 207 &  Meta & 98\\
Shanghai Jiao Tong University & 179 &  Huawei Technologies & 80\\
Nanyang Technological University & 143 & Google & 57\\
Chinese Academy of Sciences & 126 &  Baidu & 44\\
Beihang University & 124 &  Jingdong & 39\\
Carnegie Mellon University & 119 & Amazon & 21 \\
\noalign{\hrule height 1.0pt}
\end{tabular}
\label{tab:top10_institutions}
\end{table}

\begin{table}
\centering
\caption{Top 5 institutions by year (2011--2023, biennial).}
\label{tab:IAcolab_each_years}
\begin{tabular}{@{}p{2cm}p{8cm}c@{}}
\toprule
Year & Institution & \# of papers \\
\midrule
2011 & Microsoft Research Asia & 15 \\
     & Microsoft/Microsoft Research & 11 \\
     & University of Washington & 5 \\
     & Jožef Stefan Institute & 5 \\
     & Chinese Academy of Sciences & 5 \\
\midrule
2013 & Microsoft Research Asia  & 13 \\
     & Microsoft/Microsoft Research  & 10 \\
     & Tsinghua University & 5 \\
     & Tianjin University & 5 \\
     & Universitat Politècnica de Catalunya & 5 \\
\midrule
2015 & Tsinghua University & 26 \\
     & Microsoft Research Asia & 25 \\
     & Microsoft/Microsoft Research & 19 \\
     & Carnegie Mellon University & 14 \\
     & University of Science and Technology of China & 10 \\
\midrule
2017 & Microsoft Research Asia& 34 \\
     & Microsoft/Microsoft Research & 24 \\
     & Tsinghua University & 14 \\
     & Chinese Academy of Sciences & 13 \\
     & Peking University & 11 \\
\midrule
2019 & Alibaba Group & 90 \\
     & Microsoft Research Asia & 80 \\
     & Peking University & 73 \\
     & Tsinghua University & 39 \\
     & University of Science and Technology of China & 36 \\
\midrule
2021 & Alibaba Group& 137 \\
     & Microsoft Research Asia & 79 \\
     & Zhejiang University & 72 \\
     & Peking University & 45 \\
     & University of Chinese Academy of Sciences & 42 \\
\midrule
2023 & Microsoft Research Asia & 101 \\
     & Zhejiang University & 74 \\
     & Alibaba Group & 61 \\
     & University of Science and Technology of China & 57 \\
     & Tencent & 54 \\
\bottomrule
\end{tabular}
\label{tab:IAcolab_each_year}
\end{table}

Next, we analyzed the number of industry--academia collaborative papers by research institution. Table \ref{tab:top10_institutions} shows the top 10 rankings of the collaborative papers for academia and industry separately.
Among the institutions---Microsoft Research Asia and the Alibaba Group, both corporate research institutes based in China---published the most papers resulting from industry--academia partnerships. Both institutions have published more than 500 papers. They were followed by Microsoft Research which is located in the United States. On the academic side, 9 out of 10 institutions are from China, with Carnegie Mellon University being the only non-Chinese institution to rank. In contrast, the industry ranking shows more diversity, with six institutions from China and four from the United States.
\begin{figure}[p]
\includegraphics[width=0.3\textwidth,bb=9 9 358 434]{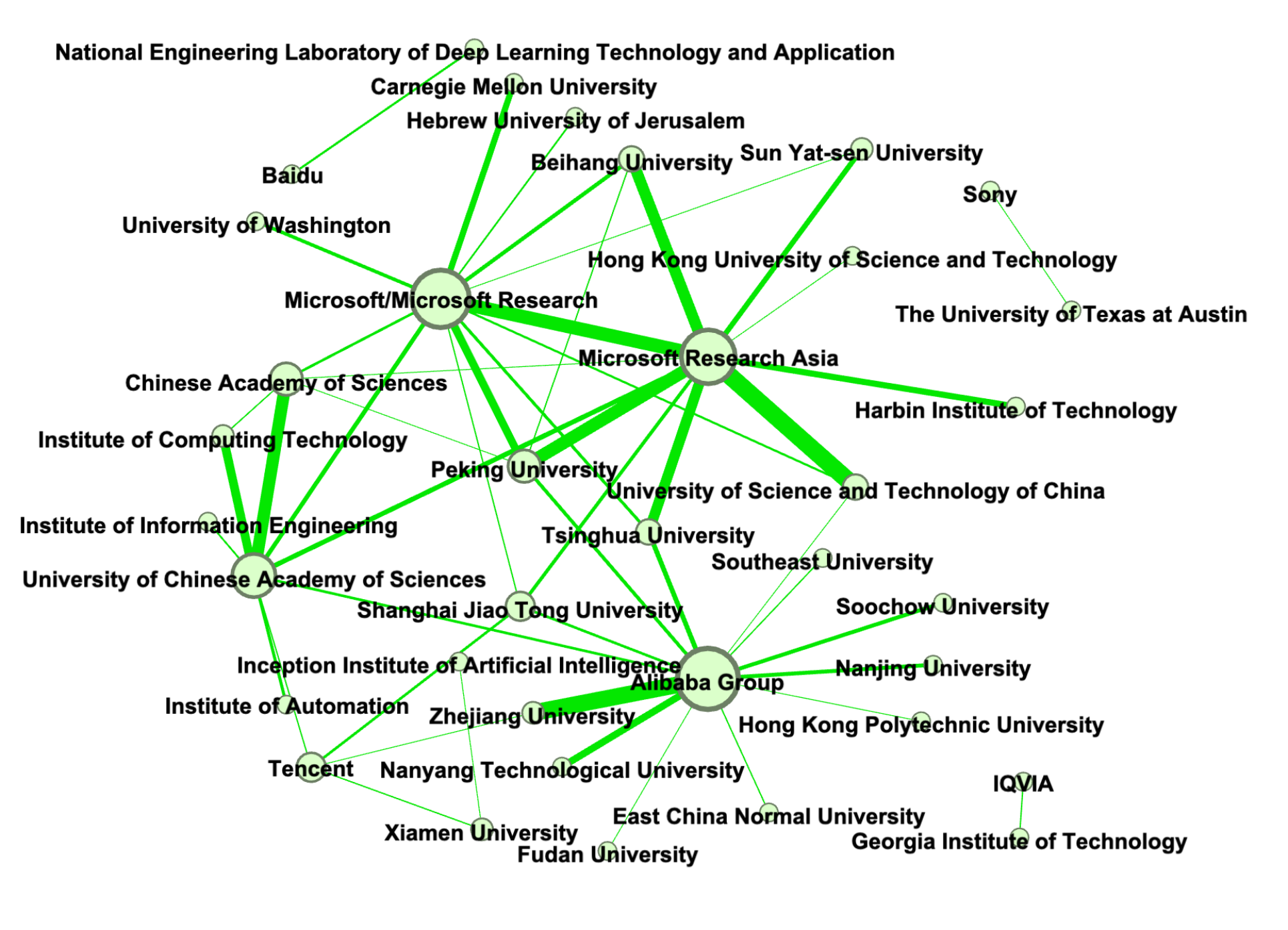}
\caption{Network of research institutions that have industry--academia collaborative papers, where edges are depicted between two institutions only if the number of their collaborative papers is greater than five.}
\label{fig:network}
\end{figure}
\begin{figure}[p]
  \begin{minipage}[b]{0.45\linewidth}
    \centering
    \includegraphics[keepaspectratio, scale=0.3, bb=9 9 358 434]{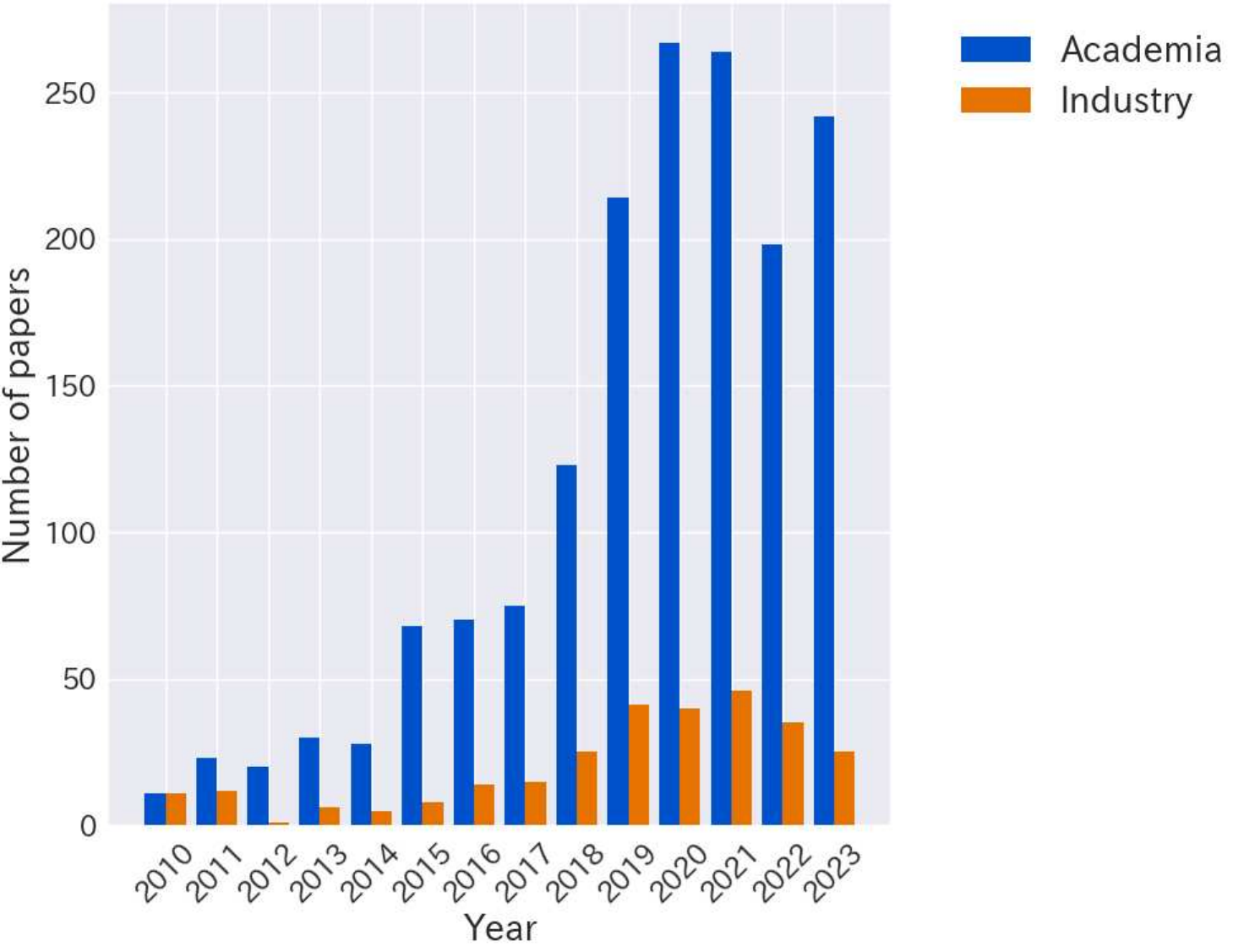}
    \subcaption{Annual differences in affiliation types of first authors in terms of the number of industry--academia collaborative papers.}
  \end{minipage}
  \begin{minipage}[b]{0.45\linewidth}
    \centering
    \includegraphics[keepaspectratio, scale=0.3, bb=9 9 358 434]{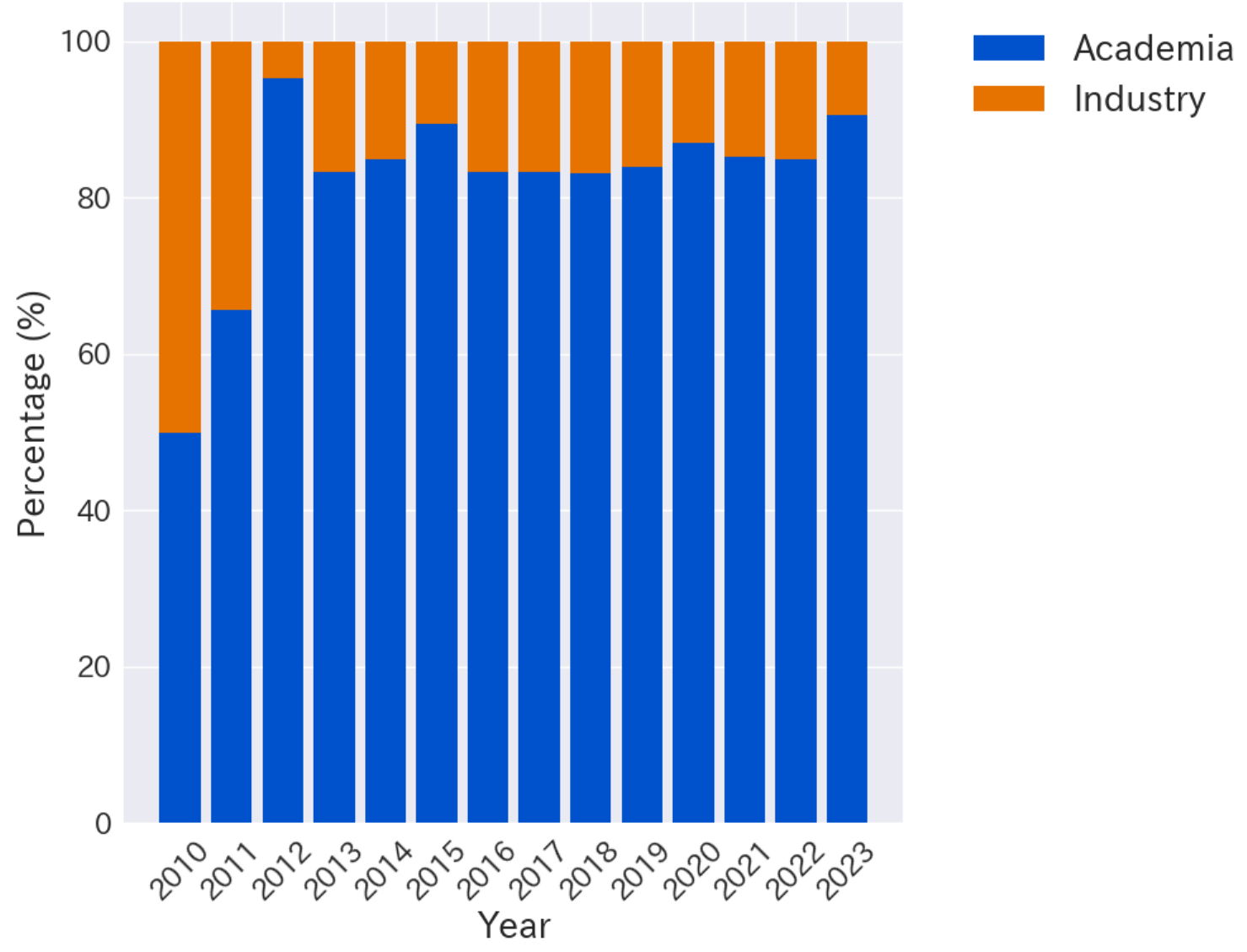}
    \subcaption{Annual differences in affiliation types of first authors in terms of the proportion of industry--academia collaborative papers.}
  \end{minipage}
  \caption{Annual differences in affiliation types of first authors in industry--academia collaborative papers.}
  \label{fig:1st}
\end{figure}
Table \ref{tab:IAcolab_each_year} shows the rankings of industry--academia collaborative papers for each year.  Microsoft Research Asia was the institution that published the most papers from 2011 to 2017, but the Alibaba Group took the lead in 2019 and 2021. In 2023, Microsoft Research Asia once again became the institution publishing the most articles. The Alibaba Group has the largest number of papers in 2019 and 2021 and the second largest number of papers in 2023. However, it is notable that the group does not appear in the annual analysis prior to 2019. This implies that collaborations with the Alibaba Group has recently increased.\par
Figure \ref{fig:network} shows the network of coauthorship relationships limited to industry--academia collaboration. The edge weights were weighted based on the number of coauthorships. Note that the Institute of Computing Technology and the Institute of Information Engineering are departments of the Chinese Academy of Sciences, but appear in this diagram because they have independent pages in the ROR. For easy viewing, we show only edges whose weights are more than five. As shown, Microsoft, Microsoft Research, and Microsoft Research Asia have strong relationships with academic institutions. In particular, Tsinghua University and Peking University are institutions with particularly high coauthorship relationships with Microsoft and its affiliated companies. However, Zhejiang University, which had the highest number of industry--academic collaborations among academic institutions, is not connected to these Microsoft entities, indicating that it does not collaborate often. The Alibaba Group, which has the second highest number of publications, frequently engages in collaborative research with Zhejiang University. Notably, companies other than Microsoft, such as Meta and Google, which have published many papers in collaboration with industry and academia, do not appear in this network.\par

This analysis reveals that Microsoft Research Asia and the Alibaba Group are the two research institutions with extremely high numbers of the collaborative papers. In addition, the academic institutions with the most collaborative papers are, in order, Zhejiang University, Tsinghua University, and Peking University—all Chinese universities. This indicates that the number of collaboration papers from China far exceeds that of other countries. In countries other than China, the number of papers from the U.S. is high. Speciﬁcally, among U.S. institutions, Carnegie Mellon University stands out for academic contributions, whereas companies such as Microsoft (including Microsoft Research), Meta, and Google are notable for their high number of industry collaborations in academic papers. Although Zhou et al. \cite{Zhou2016-ay} showed that China lags behind the United States in terms of productivity and collaboration intensity in industry--academia collaboration, the results of focusing on AI conferences seem to suggest the opposite trend.

\subsection{Analysis of First Author Affiliations in Industry--Academia Collaborative Papers}

In the AI research field, the first author is typically responsible for composing the paper. By analyzing the affiliations of the first author of an industry--academia collaboration paper, it is possible to ascertain which organization has the initiative in publishing the research results as a paper. Figure \ref{fig:1st} shows the time series change in the number and proportion of papers based on the affiliation type of the first author in the collaborative papers. Here, although the number of papers increased from 2010 to 2017, the number of papers with first authors from industry barely increased. Starting from 2017 onward, there was an increasing trend until 2021, but it has been decreasing since then. When it comes to proportions, the ratio of academia to industry was exactly half and half in 2010, the starting year of the analysis. However, after academia reached its highest proportion in 2012, it has remained constant at approximately 80\%. This indicates that industry--academia partnerships result in a paucity of papers led by industry, with the majority of AI research papers being authored by academic institutions.

\section{Content-Based Analysis}
\begin{table}[t]
\centering
\caption{Results of binary classification using SciBERT to determine industry--academia collaborative papers.}
\begin{tabular}{l|c|c|c|c}
\noalign{\hrule height 1.0pt}
Method & Accuracy & Precision & Recall & F1-score \\
\noalign{\hrule height 1.0pt}
With negative sampling (Random) & 0.49 & 0.49 & 0.49 & 0.49 \\
With negative sampling (SciBERT) & 0.61 & 0.62 & 0.61 & 0.60 \\
Without negative sampling (Majority) & - & 0.46 & 0.50 & 0.48 \\
Without negative sampling (SciBERT) & - & 0.58 & 0.51 & 0.49 \\
\noalign{\hrule height 1.0pt}
\end{tabular}
\label{tab:bert}
\end{table}
We analyzed the differences in content using the abstracts of industry--academia collaborative papers. To do so, we first extracted the abstracts from all PDFs via GROBID. Then, we trained a binary classifier that classified a given abstract into the collaborative work or not. A high classification performance would mean that the content can be distinguished in terms of textual features used by the classifier, whereas showing a low classification performance would imply that the classifier cannot determined the differences in content. Specifically, we fine-tuned a pretrained SciBERT \cite{beltagy-etal-2019-scibert} model with the collaborative papers as positive examples and papers not involving companies as negative examples. This was a classification of imbalanced data with 1,919 the collaborative papers and 18,353 papers not involving companies. We split the dataset into training, validation, and test sets at a ratio of $8:1:1$, ensuring a consistent distribution of collaborative and noncollaborative papers across all sets. We performed classification both on the imbalanced data as is and after applying negative sampling. Table~\ref{tab:bert} shows the results. The top two rows show the results of classification on the imbalanced data without negative sampling. The initial row denotes the random classification result, which serves as the baseline. The bottom two rows show the results after applying negative sampling to match the number of noncollaborative papers to the number of collaborative papers. The third row from the top represents the baseline result, which forcibly classifies all data into the majority class (i.e., noncolaborative papers). Regarding the results without negative sampling, the F1-score is at 0.49 which is the same as that of random classification, suggesting that classification was not possible. For the results with negative sampling, they were  just barely above 0.6 for all evaluation indicators, indicating that the classification was not very successful. These classification results suggest that there may not be much of a difference in content between the collaborative papers and other papers.\par
Comparing this result with the findings of these studies \cite{farber2024analyzing,Jee2022-gx}, it appears that, although the content of research does not differ from that of papers involving only academia, the impact is different. This is an interesting result, and as we consider that complex factors are involved, we are planning a more detailed analysis.

\section{Limitations}
There are several limitations in the current study. First, although GROBID is an excellent bibliographic information extraction library, omissions and leaks can occur, even when it is carefully used in Section 3. Our results might be aﬀected by these mistakes. The accuracy of S2AFF is also a limitation. Although the GitHub page of S2AFF\footnote{https://github.com/allenai/S2AFF} shows very high performance, with a recall of 0.983, this performance is for a very well-curated dataset and not for strings extracted by GROBID.

\section{Conclusion}
The present study employed bibliometric analysis to examine industry--academia collaborative papers in the ﬁeld of AI, with a particular focus on two top-tier conferences. The papers presented at these conferences were collected, and the bibliographic information was directly extracted. The analysis yielded answers to each of the four research questions. In response to the initial research question, the number of collaborations in the ﬁeld of AI demonstrated an upward trajectory from approximately 2017 to 2020, reaching a peak around 2020. During this period, academic institutions that had previously engaged in few such collaborative endeavors began to increase the number of papers produced. Regarding the second research question, the majority of research institutions that were actively engaged in industry--academia collaboration were located in China. Of these, the research institutions with the greatest number of industry--academia collaborations were Microsoft Research Asia and the Alibaba Group from industry and Zhejiang University and Tsinghua University from academia. Regarding countries other than China, numerous research institutions in the United States have engaged in collaborative research, with Microsoft and Carnegie Mellon University being particularly active here, as evidenced by the large number of papers they produced together. Third, the ﬁrst authors of the collaborative papers were predominantly aﬃliated with academic institutions, indicating that academia remains the primary driver of writing papers, even in collaborative eﬀorts involving industry. Finally, an analysis of the content of the papers revealed minimal diﬀerences between those that were collaborative and those that were solely academic. Although industry involvement is anticipated to meet practical corporate requirements from the industrial sector, our ﬁndings indicate that this may not be substantially reﬂected in the research content. However, this result may involve complex causal relationships, thus warranting further investigation. Future research directions include improving the performance of models such as GROBID and S2AFF and extending the analysis of the collaboration to broader ﬁelds within and beyond computer science.
\bibliographystyle{splncs04}
\bibliography{mybibliography}
%




\end{document}